\documentclass{aastex62}
\usepackage{color} 
\usepackage{tabularx} 
\usepackage{epsfig}
\usepackage{amsmath} 
\usepackage{amssymb} 
\usepackage{bm}
\usepackage{graphicx}
\usepackage{wasysym}
\usepackage{multirow}
\usepackage{threeparttable}
\definecolor{purp}{HTML}{8904B1}
\newcommand{\rsun}{R$_\odot$}

\shorttitle{Changes in solar structure}
\shortauthors{Watson and Basu}

\begin{document}

\title{Solar Cycle Related Changes in the Helium Ionization Zones of the Sun}

\author[0000-0001-8456-4142]{Courtney B. Watson}
\affil{Department of Astronomy, Boston University, Boston, MA 02215, USA}
\affil{Department of Astronomy, Yale University, New Haven, CT 06520, USA}
\email{cbwatson@bu.edu}

\author[0000-0002-6163-3472]{Sarbani Basu}
\affil{Department of Astronomy, Yale University, New Haven, CT 06520, USA}

\begin{abstract}
Helioseismic data for solar cycles 23 and 24 have shown unequivocally that solar dynamics changes with solar activity. Changes in solar structure have been more difficult to detect. Basu \& Mandel (2004) had claimed that the then available data revealed changes in the He{\sc ii} ionization zone of the Sun. 
The amount of change, however, indicated the need for larger than expected changes in the magnetic fields. Now that helioseismic data spanning two solar cycles are available, we have redone the analysis using improved fitting techniques. We find that there is indeed a change  {in the region around the} He{\sc ii} ionization zone that is correlated with activity. Since the data sets now cover two solar cycles, the time variation is easily discernible. 
\end{abstract}

\keywords{Solar Physics: Helioseismology --- Solar Physics: solar activity --- Solar physics: solar interior --- Stellar Interiors}

\section{Introduction}

It has been known for about thirty-five years that solar oscillation frequencies increase with activity \citep{woodard1985, yvonne, woodard}.  {Subsequent analyses showed that the frequency changes were correlated with the average line-of-sight magnetic field over the solar surface \citep{woodard1991ApJ}, and this was confirmed later using much longer data sets \citep[e.g.][]{howe2002}. The correlation between frequency shifts and activity is also seen for other solar activity proxies \citep[etc.]{kiran,howe2018}}. Both ground-based and space-based data have shown that solar dynamics changes dramatically over the solar cycle \citep[see e.g.,][etc]{howe_review,basu2019}. Changes in solar structure, however, are more difficult to determine. There have been some indications of changes in structure at the base of the convection zone (CZ) \citep{effdarwich, baldner} and indications are that changes are very small. There is also evidence that the immediate sub-surface layers of the Sun change with activity and that there is a latitudinal dependence of the changes\citep{antia2001, basu2007}. 

\citet{basu2004}, henceforth BM04, examined the then available data from the Global Oscillation Network Group (GONG) \citep{hill1996} and the Michelson Doppler Imager (MDI) on board the Solar and Heliospheric Observatory spacecraft \citep{mdi} to look for changes slightly deeper into the CZ, specifically the second helium ionization zone around 0.98\rsun\ based on work by \citet{gough2002} who suggested that there could be solar-cycle related variations in that region. The helium ionization zone is an ``acoustic glitch,'' a region where the sound speed changes rapidly over a region shorter than the radial wavelength of the modes. A second, sharper glitch is the base of the CZ where the glitch is caused because of the transition of the temperature gradient from being radiative in the deeper layers to adiabatic in the CZ. In the ionization zone, the glitch is caused by the decrease in the adiabatic index, $\Gamma_1$, because of ionization. An acoustic glitch contributes a characteristic oscillatory signature in the frequencies \citep{gough1990}. The wavelength of the oscillations is the acoustic depth of the glitch while the amplitude of the oscillations is governed by how ``severe" the glitch is. In the case of the He{\sc II} ionization zone, normally the amplitude would be a measure of the helium abundance --- the higher the abundance of helium the higher the amplitude. A change in $\Gamma_1$ can also occur in the presence of magnetic fields \citep{basuetal2004} because of changes in the local temperature. Thus a changing magnetic field in the  {in the layers around the} He{\sc II} ionization zone will change the amplitude of the signal. BM04 measured the amplitude and found that it decreases as the flux of the 10.7 cm flux, an activity indicator, increases. 

The BM04 results were confirmed by \citet{verner} and \citet{Ballot2006} using low-degree data from the Birmingham Solar Oscillation Network (BiSON) and the Global Oscillations at Low Frequencies (GOLF) respectively. BM04 interpreted the change in the ionization zone as one caused by a change in temperature. These results, and those of \citet{verner} and \citet{Ballot2006}, were however challenged by \citet{Gough2013}. \citet{Gough2013} analyzed the potential significance of the results under the assumption that the change in the amplitude of the  He{\sc ii} signal is caused by a dilution of the influence of the ionization zone on the wave propagation speed by a broadly distributed magnetic field. He concluded that if the variations of the He{\sc ii} signal were a direct result of the presence of a temporally varying large-scale magnetic field, then the total solar cycle change of the spatial average of the magnetic field $<B^2>$ in the vicinity of the He{\sc ii} ionization zone is greater than most estimates. 

The BM04 analysis work was done with data for about half a solar cycle. We now have data for two solar cycles. In this work, we repeat the work of BM04 but use improved fitting methods. BM04 had not taken correlated errors into account, we do that now. Additionally, the previous work had assumed that the uncertainty on the result of every set is the same; we abandon that assumption and determine uncertainties for each set by boot-strapping. Another change we make is that we use two different methods to fit the data --- the model is highly non-linear and there is often a tendency for the solution to be stuck at a local $\chi^2$ minimum.

\section{Data, model, and fitting techniques\label{sec:data}}

We use helioseismic data from three sources: (I) The ground-based Global
Oscillation Network Group (GONG) \citep{hill1996}, (II) the Michelson Doppler Imager (MDI) on board the Solar and Heliospheric Observatory spacecraft \citep{mdi} and (III) the Helioseismic and Magnetic Imager (HMI) \citep{hmi} on board the Solar Dynamics Observatory. 

We use 236 GONG data sets that cover a period from 1995 May 7 to 2018 March 18. The data are labeled by GONG ``months'', each ``month'' being 36 days long. Solar oscillation frequencies and splittings of sets starting Month~2 are obtained using 108 day (i.e., 3 GONG months) time series. There is an overlap of 72 days between different data sets. Data from MDI cover the period from May 1, 1996 to April 24, 2011, with frequencies in each set obtained from 72 day time series; the sets have no overlap in time. We use 48 sets of HMI data with the first set starting at 2010 April 30 and the last ending on 2019 October 16. Like MDI, HMI frequencies sets are obtained with non-overlapping 72-day time series. Note that we essentially use data covering two full solar cycles. We use the 10.7 cm radio flux as an indicator of solar activity (\citealt{tapping2} and \citealt{tapping})\footnote{Available from
 http://www.spaceweather.gc.ca/solarflux/sx-en.php, 
https://omniweb.gsfc.nasa.gov/form/dx1.html}

In order to compare our results with those of BM04, we use the same model to determine the amplitude of the glitch signature. As with that work, we first amplify the oscillatory signature by taking the fourth-differences of the frequencies:
\begin{equation}
\delta^4\nu_{n,\ell}=\nu_{n+2,\ell}-4\nu_{n+1,\ell}+6\nu_{n,\ell}
-4\nu_{n-1,\ell}+\nu_{n-2,\ell},
\label{eq:fourth}
\end{equation}
where $\nu_{n,\ell}$ is the frequency of a mode of radial order $n$ and degree $\ell$.
The fourth differences are fitted to the form 
\begin{eqnarray}
\delta^{4}\nu
&= \left[a_1+ a_2\nu +{(a_3+a_4L)/\nu^2}\right]\; +\nonumber\\
&  \left(a_5+\frac{a_6}{\nu_m^2}
 +\frac{a_7L}{\nu_m^2}+\frac{a_8L}{\nu_m^4}\right)\sin(2\nu_m\tau_a
 +\phi_a)\; + \nonumber\\
&
\left(b_1+\frac{b_2}{\nu_n^2}+\frac{b_3L}{\nu_n^2}+\frac{b_4L}{\nu_n^4}\right)
\sin(2\nu_n\tau_b+\phi_b)\label{eq:model}
\end{eqnarray}
where
\begin{equation}
L=\ell(\ell+1), \quad
\nu_m=\nu-{{\gamma_aL}/{2\tau_a\nu}},\hbox{ and}\quad
\nu_n=\nu-{{\gamma_bL}/{2\tau_b\nu}}.
\label{eq:deff}
\end{equation}
The parameters $a_1$--$a_4$ define the overall smooth term, $a_5$--$a_8$, $\tau_a$ and $\phi_a$ define the oscillatory component due to the He{\sc ii} ionization zone, and the terms $b_1$--$b_4$, $\tau_b$, and $\phi_b$ define the oscillatory component due to the base of the convection zone. The frequencies of the two components, $\tau_a$ and $\tau_b$, are approximately the acoustic depths of the He{\sc ii} ionization zone and the CZ base respectively; they also include a contribution from the frequency-dependent part of the phases $\phi_a$ and $\phi_b$ that is not taken into account explicitly.  {The terms $\gamma_a$ and $\gamma_b$ take into account the fact that the response to acoustic glitches has a degree dependence, and these two parameters are fitted along with the other parameters. The shifted frequencies 
$\nu_m$ and $\nu_n$ in Eq.~\ref{eq:deff} are needed to account for the degree-dependence of the response to acoustic glitches; the derivation of this term {can be found} in \citet{overshoot}. }

The parameters are obtained with non-linear minimum-$\chi^2$ fits, and this is where we differ from BM04. Eq.~\ref{eq:fourth} shows that the errors of neighboring fourth differences will be correlated. BM04 did not take error-correlations into account and assumed that the diagonal terms of the full error-covariance matrix sufficed. In this work, we use the full error covariance matrix, and hence, we define $\chi^2$ as
\begin{equation}
\chi^2 = (y - y')^{T} C^{-1}(y-y'),
\label{eq:chi2}
\end{equation}
where $y$ is a column matrix with the data, $y'$ the result of the model fit, and $C$ is the covariance matrix.

Like BM04, we fit the form in Eq.~\ref{eq:model} in two degree ranges $5 \le \ell \le 25$, and $30 \le\ell\le 60$.  {All modes in the first set have lower turning points below the CZ base. For the second set, we choose only those modes that have lower turning points above the CZ base}. Modes in the first set (the 'low-degree' set) sample both glitches, the He{\sc II} ionization zone, as well as the CZ base and consequently all three terms of Eq.~\ref{eq:fourth} are fitted. Modes in the second set (the `high-degree' set) sample only the He{\sc II} ionization zone and consequently only the first two terms of Eq.~\ref{eq:model} apply. Given that the amplitude of the two oscillatory terms in Eq.~\ref{eq:model} are a function of frequency, like BM04, we use an average amplitude between 2 and 3.5 mHz for our investigation.  {We define the average amplitude $<A>$ such that it does not depend on the degree. Defining $\nu_{\rm min}$ and $\nu_{\rm max}$ to be the minimum and maximum frequency, we define $<A>$ as
\begin{equation}
<A>=\frac{\int_{\nu_{\rm min}}^{\nu_{\rm min}} \left(a_5 + \frac{a_6}{\nu^2}\right)d\nu}{\int_{\nu_{\rm min}}^{\nu_{\rm min}}d\nu}
= \frac{1}{\nu_{\rm max}-\nu_{\rm min}} \left(a_5(\nu_{\rm max}-\nu_{\rm min})+a_6\left[\frac{1}{\nu_{\rm max}} -\frac{1}{\nu_{\rm min}}\right]\right).
\label{eq:ave}
\end{equation}
}

We use two different codes to do the fits. The first is the same as the one used by BM04 which uses the Broyden--Fletcher--Goldfarb--Shanno (BFGS) algorithm as implemented by \citet{bfgs}. The second is one that uses the python code \verb|scipy.curve| which uses the Levenberg-Marquardt (LM) algorithm to fit the data. The highly non-linear nature of the model means that the $\chi^2$ surface has many local minima, given the differences between the two algorithms, a similar result returned by the two different methods is unlikely to be that from a local minimum. The complicated model also means that whether or not we obtain a successful fit within the stipulated number of iteration depends on the initial guesses. In another departure from BM04, we have increased the number of initial guesses in order to obtain 250 sets of fitted parameters to find a global minimum. However, we go even further and determine the uncertainties using a traditional boot-strapping method by simulating 100 realizations of the observations, fitting each one of them in the same manner as the original data. We use the aggregate of results to determine the value of the mean amplitudes and their uncertainties.

\begin{figure}
\epsscale{0.75}
\plotone{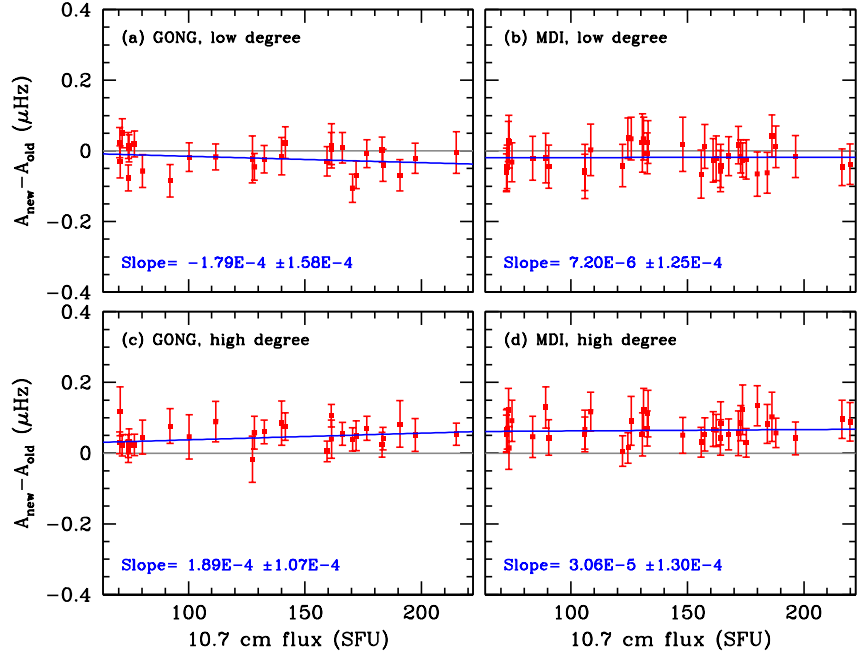}
\caption{ The difference between the average amplitude of the glitch signature between 2 and 3.5~mHz obtained in this work and the work of BM04 that did not take the complete error-correlation matrix into account. The gray line shows the no-difference case, while the blue line is a weighted least-squares fit to the differences.  Note that for GONG data the differences show a slope with respect to the 10.7 cm flux in solar flux units (1~SFU~$= 10^4$ Jansky $=10^{-22}$ W~m$^{-2}$~Hz$^{-1}$). This indicates that the slope obtained in this work will be different from those obtained earlier even if we restricted ourselves to the same data set. For MDI, there seems to be a constant difference, rather than a difference in slope.
}
\label{fig:comp}
\end{figure}

We first compare the amplitudes obtained by BM04 with the new fits with the same code but using the full error-correlation matrix, these can be seen in Fig.~\ref{fig:comp}. It should be noted that BM04 only fitted non-overlapping sets of GONG data and there were no HMI data at that time. Note that while there are small differences in the result obtained --- the differences are at most 1.5$\sigma$, there are systematic differences. In the case of GONG, there is a slope with respect to the 10.7 cm flux which implies that any correlation that we obtain will be somewhat different than that found by BM04. For MDI, the differences appear to be a constant. For both GONG and MDI, the new amplitudes are higher than the old ones for the higher-degree cases.

\begin{figure*}
\epsscale{0.75}
\plotone{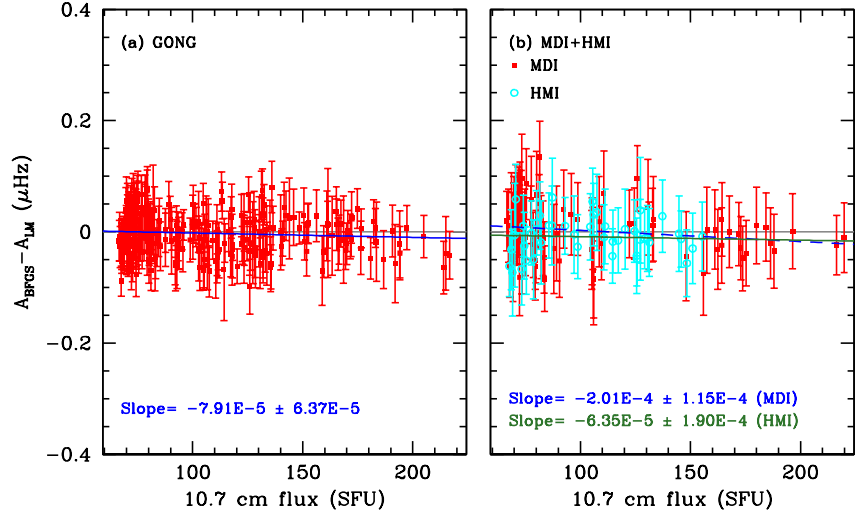}
\caption{ The difference between the average amplitude of the glitch signature between 2 and 3.5~mHz obtained between the BFGS and LM codes for the low-degree mode set for (a) GONG and (b) MDI and HMI data. The grey line in each panel marks the no-difference case. There is a small, but statistically insignificant differences in the slopes with respect to the 10.7 cm flux obtained with GONG data (blue line in Panel~a) and HMI data (green line in Panel~b). The slope for the MDI sets is slightly more significant. Note that if MDI and HMI are used together, as we do later, then the differences have a slope of $1.64  \times 10^{-4}\pm 9.61\times 10^{-5}$.
}
\label{fig:comp2}
\end{figure*}

In Fig.~\ref{fig:comp2} we show the difference in results obtained with the BFGS code and the LM code. We were unable to get stable fits with the LM code for the high-degree set sets. The number of converged results were too low to estimate the amplitude and their uncertainties. We are not completely sure why this is the case; the solutions for most parts seemed to get stuck in local minima and often the reduced $\chi^2$ was greater than 100.  Consequently, we concentrate on results obtained for the low-degree set. The results for the GONG sets obtained by the two codes are not very different, neither are those for the HMI sets. The difference in results for the MDI sets on the other hand show a somewhat significant slope with respect to the 10.7 cm flux. However, the quality of the LM fits to the space-based data are worse than the GONG case, most likely because each GONG set is obtained with 105 days of data, while MDI and HMI sets were obtained with only 72 days of data.  The difference in the results obtained by the two codes thus has to be taken into account while judging if there is a change in the amplitude of the helium glitch signature with change in solar activity.

\section{Results \label{sec:results}}

\begin{figure*}
\epsscale{0.85}
\plotone{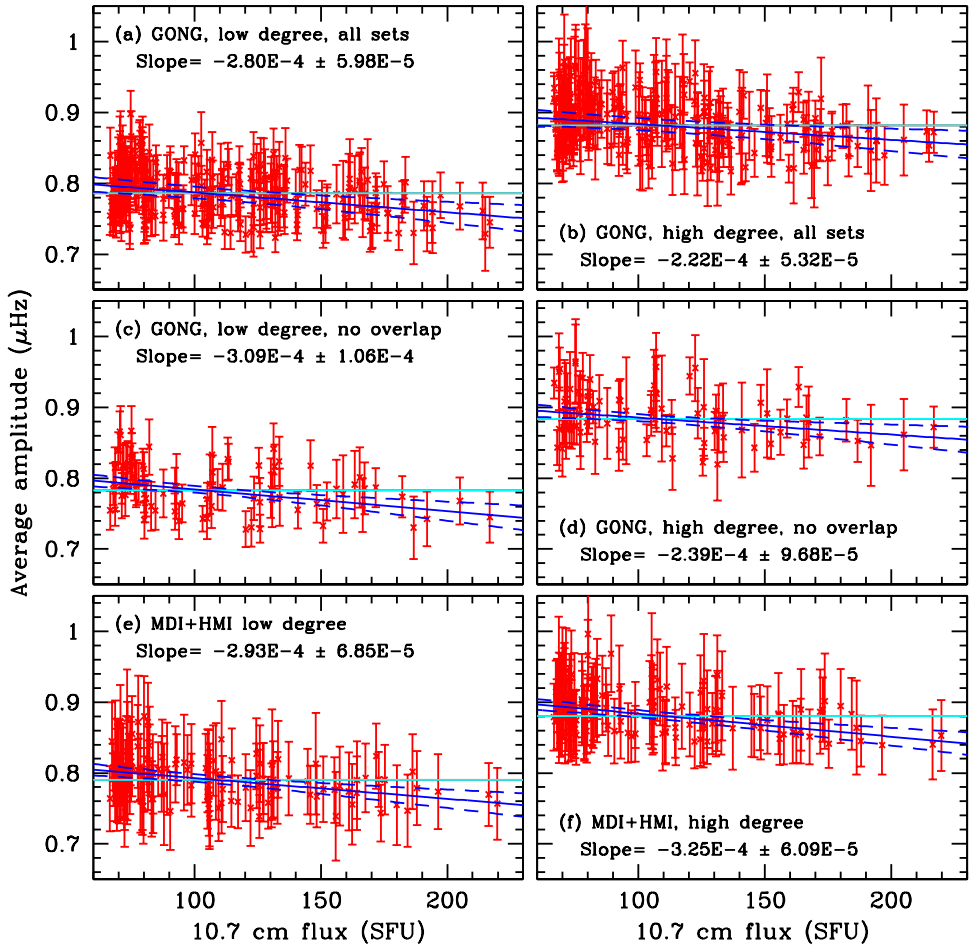}
\caption{ The average amplitude of the helium glitch signature plotted as a function of the 10.7 cm flux. The results were obtained with the BFGS code. In each panel, the cyan horizontal like is the weighted average of the points. The blue solid line is a weighted least-squares fit to the points with the 95\% confidence limit plotted as the blue dashed lines.  Panels (a), (c), and (e) are low-degree results, while panels (b), (d), and (f) show high-degree results. We show all the sets for GONG in panels (a) and (b) and only non-overlapping sets in panels (c) and (d). 
}
\label{fig:all}
\end{figure*}

The average amplitude of the glitch signature is plotted as a function of the 10.7 cm flux in Fig.~\ref{fig:all}. The results shown are for the BFGS code. In all panels, we see that that the amplitude decreases when the activity index increases. The slope is much larger than those caused by the changes caused by using the full error-correlation matrix. The slope of the change as given by the LM fits to the low-degree sets corresponding to panels (a), (c), and (e) are, respectively,
$-2.71  \times 10^{  -4}\pm 4.67  \times 10^{  -5}$, $-2.88  \times 10^{  -4}\pm 8.42  \times 10^{  -5}$ and $-2.23  \times 10^{  -4}\pm 9.61  \times 10^{  -5}$. Thus, the slopes are less steep than those returned by the BFGS code; nevertheless, they are significant. We therefore believe that, although the exact correlation between the glitch amplitude the 10.7 cm flux is uncertain, the trend is robust. 

\begin{figure}
\epsscale{0.65}
\plotone{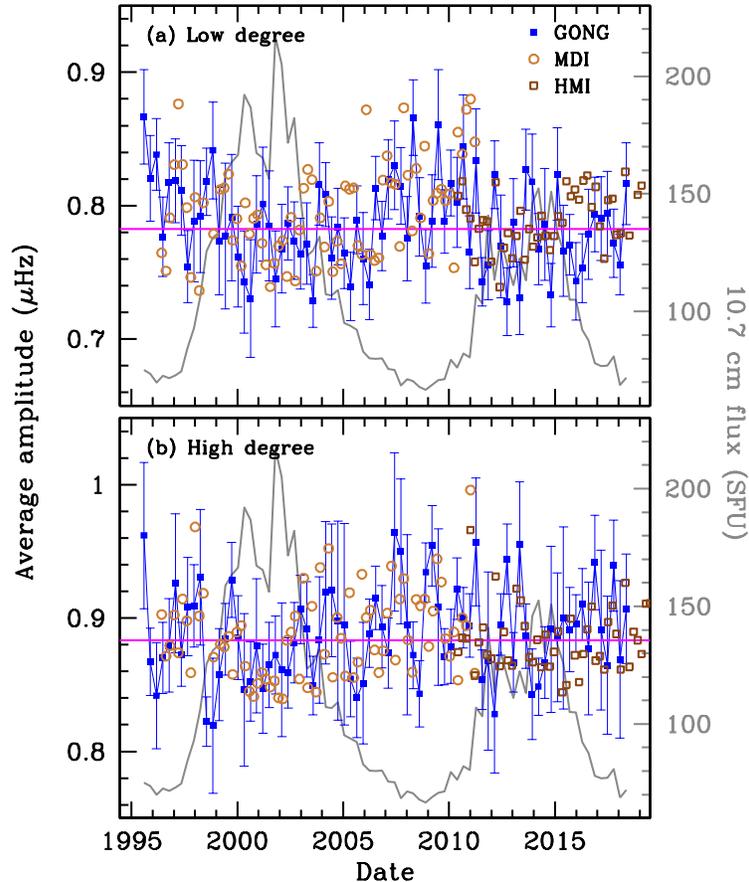}
\caption{ The average amplitude of the He glitch signature plotted as a function of time (points) for (a) low-degree and (b) high-degree sets. The magenta line is a weighted average of the blue points which show results for non-overlapping GONG sets.  {The results are from the BFGS codes. } The grey curve in the background is the 10.7 cm flux whose value can be read from the axis on the right. 
}
\label{fig:time}
\end{figure}

The two solar cycles' worth of data allows us to plot the results against time to examine whether or not a change with solar activity is easily discernible. We show this in Fig.~\ref{fig:time}. Note that the time-variation and the anti-correlation with solar activity is very clear for the low-degree sets, though visually less clear for the high-degree ones.

\section{Discussion and Conclusions \label{sec:conc}}

We have repeated the work of BM04 with data for two complete solar cycles to determine whether or not there are changes inside the Sun at depths corresponding to the He{\sc ii} ionization zone. This was done by measuring the amplitude of the signature of the acoustic glitch caused by the second ionization of helium. We find that the amplitude of the glitch indeed changes with time and that it is anti-correlated with solar activity. 
This behavior is seen in both ground-based and space-based data. This shows, unambiguously, that there are solar-cycle related changes in the Sun at these depths. 

\begin{figure}
\epsscale{0.65}
\plotone{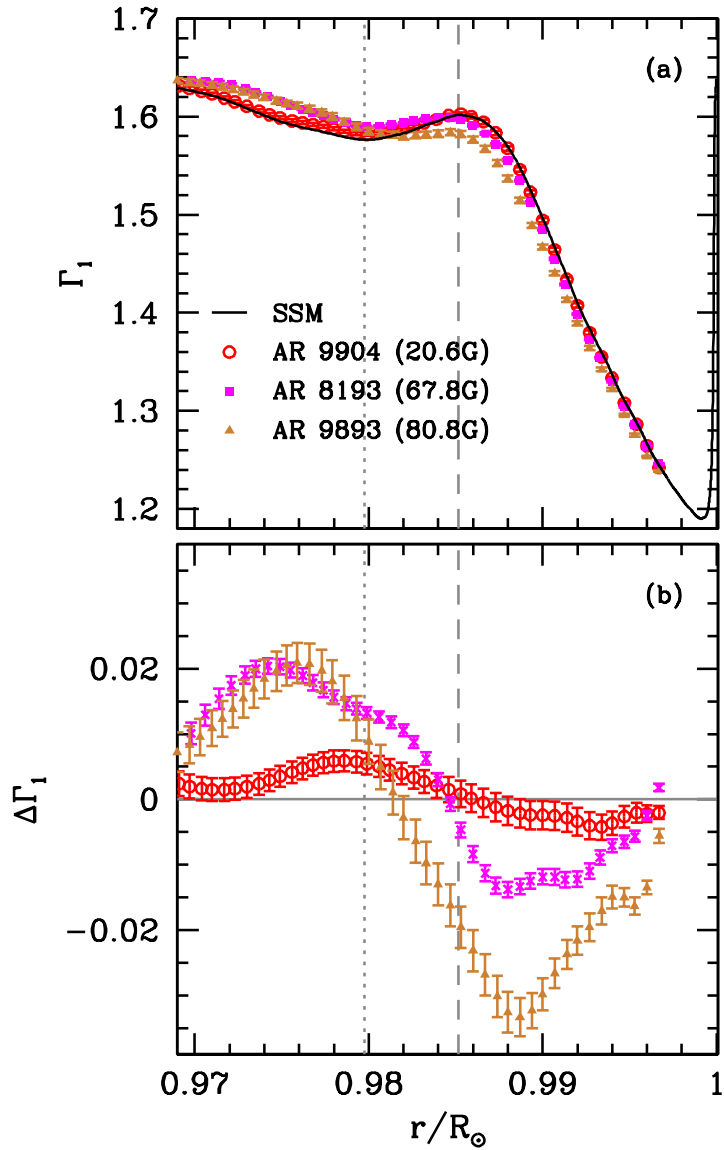}
\caption{ (a) The $\Gamma_1$ profile of a standard solar model (SSM) and of the region below three active regions. (b) The $\Gamma_1$ difference between the active regions and the quiet Sun. The active region inversion results are from \citet{basuetal2004}. The magnetic activity index defining the magnetic field of the region studied is also indicated for each active region. The dotted vertical line marks the position of the minimum in $\Gamma_1$ in the He {\sc ii} ionization zone for the model, the dashed line marks the peak in $\Gamma_1$ between the He{\sc i} and He {\sc ii} ionization zones. 
}
\label{fig:gamma}
\end{figure}

 {The results, which are very similar to those obtained by BM04, lead to the question of why the \citet{Gough2013} analysis led to magnetic-field estimates {of nearly 300 Kilo Gauss, an unrealistically large field}. We believe there are two reasons for this. The first is that the analysis was based on the common assumption that the oscillatory signature in the frequencies arises solely from the dip in $\Gamma_1$ caused by He{\sc ii} ionization. Work that has been done subsequently \citep{broomhall, kuldeep} however, showed that the signature arises from a depth that corresponds to the peak in $\Gamma_1$ between the He{\sc i} and He{\sc ii} ionization zones (shown in Fig.~\ref{fig:gamma} as the vertical dashed line). Shallower layers have lower gas pressure and hence need smaller magnetic fields to change their structure. 
{Direct inversion results of solar-cycle related changes in global modes are too noisy to determine how changes in solar magnetic activity modify the $\Gamma_1$ profile, consequently} we look instead to $\Gamma_1$ changes below sunspots of different strengths and these are shown in Fig.~\ref{fig:gamma}. {These results were obtained by taking  $15^\circ \times 15^\circ$ regions around sunspots. Using such a large region considerably dilutes their magnetic field, but also allows one to obtain results up to depths corresponding to the He{\sc ii} ionization zone. The magnetic field of this large region is therefore lower than that of the active region itself. 
Inversion results for three such regions with surface magnetic field strengths of about 20, 68, and 80 Gauss are shown in the figure. The lower panel shows that a larger change takes place in the boundary between the He{\sc i} and He{\sc ii} ionization zones. The second reason is that the \citet{Gough2013} analysis, based on \citet{houdek}, assumed that the He{\sc ii} can be modeled as a Gaussian with a finite width and that the magnetic fields change the amplitude of the Gaussian. However, as can be seen from Fig.~5, while magnetic fields do make the He {\sc ii} ionization-related dip in $\Gamma_1$ smaller, they also distort the shape of the dip. Thus the model used to associate the changes in amplitude to changes in $\Gamma_1$, and then the conversion of the change in $\Gamma_1$ needs to be changed. All these factors need to be accounted for to determine a change in the magnetic fields from the amplitude of the glitch signature.}} 

 {A detailed analysis carried out by \citet{baldner} revealed that between solar Cycle~23 maximum and minimum, there were small but significant changes in the sound speed at the equator and at a latitude of $15^\circ$. The results were too noisy at higher latitudes to see anything, but at the low latitudes the changes were seen even below $0.9$\rsun. Our results indicate that changes also occur in a spherically symmetric sense.  }

\acknowledgements  {We would like to thank the referee for comments that have helped us improve the paper.}


\end{document}